# Gradients of Connectivity as Graph Fourier Bases of Brain Activity

Giulia Lioi[1,*], Vincent Gripon[1], Abdelbasset Brahim[2], François Rousseau[1], and Nicolas Farrugia[1]

[1]IMT Atlantique, Brest, France
[2]INSERM, Laboratoire Traitement du Signal et de l'Image (LTSI) U1099, University of Rennes, Rennes, France
[*]corresponding author: Giulia Lioi (giulia.lioi@imt-atlantique.fr)

## ABSTRACT

The application of graph theory to model the complex structure and function of the brain has shed new light on its organization and function, prompting the emergence of network neuroscience. Despite the tremendous progress that has been achieved in this field, still relatively few methods exploit the topology of brain networks to analyze brain activity. Recent attempts in this direction have leveraged on graph spectral analysis and graph signal processing to decompose brain activity in connectivity eigenmodes or gradients. If results are promising in terms of interpretability and functional relevance, methodologies and terminology are sometimes confusing. The goals of this paper are twofold. First, we summarize recent contributions related to connectivity gradients and graph signal processing, and attempt a clarification of the terminology and methods used in the field, while pointing out current methodological limitations. Second, we discuss the perspective that the functional relevance of connectivity gradients could be fruitfully exploited by considering them as graph Fourier bases of brain activity.

## 1 Introduction

Modern attempts at understanding brain function have leveraged the use of graph theory to grasp complex properties of neuronal networks, giving rise to the field of network neuroscience[1,2]. Modeling brain organization using graphs has led to fascinating results, such as the brain's hypothetical rich-club organization[3], the cortical organization in functionally relevant modules[4], as well as common wiring principles across species[5]. Despite the tremendous progress that has been achieved in network neuroscience, surprisingly relatively few methods such as graph signal processing (GSP)[6] exploit brain connectivity (i.e. take into account the topology of brain network) to characterize brain activity.

First steps in the direction of exploiting connectivity graphs in the analysis of brain signals have been made using spectral graph theory[7,8]. The underlying idea of this latter is to interpret the eigenvectors of graph Laplacians as harmonic components. Increasing evidence of the functional relevance of these spectral components of brain networks (i.e. *connectivity gradients*, eigenmodes or harmonics[9–11], see Table 1), has recently been shown with a variety of approaches [12–14].

GSP takes a further step towards a joint analysis of brain activity and structure. The emergence of GSP is mostly due to the elegant and powerful analogy between graph Laplacian eigenvectors and classical Fourier analysis [15] and the possibility to decompose a signal "living on graph" as a combination of spatial harmonics. Recent works have exploited GSP to decompose brain activity in graph Fourier modes, with encouraging results[16–19].

The gradients/GSP framework is complementary to the classical approach of mapping cortical areas functions (brain parcellations) and discrete networks associated with a particular condition or task. Exploiting the topology of brain networks, brain activity or structure is decomposed as a *continuum* of graph spectral components. This novel approach has been successfully applied to the analysis of the healthy or pathological brain in an increasing number of studies. However, readers that first approach the recent literature are often confronted with different notations, terminology and methods (i.e. Laplacian embedding, diffusion maps, graph Fourier modes, see Table 2) that sometimes are not consistent between the GSP or "gradients" communities and may be difficult to unravel.

The goal of this paper is first to summarize recent contributions that have used connectivity gradients and GSP for neuroimaging. In particular, we aim at clarifying terminology, compare different methodologies, provide resources (see Box 2) and key references (Tables 1 and 2) to bring some order into the field. The second goal of this paper is to make a link between the two frameworks by discussing Gradients as Graph Fourier Bases. We argue that using GSP for the analysis of multimodal neuroimaging data will pave the way to more interpretable analysis methods.

## 2 Gradients of Brain Connectivity

Connectivity gradients are obtained from graph spectral decomposition of the connectivity matrix. As described in box 1, they correspond to Laplacian eigenvectors **u** of connectivity graphs. Connectivity gradients provide a representation of cortical organization as a continuum of spatial harmonics that can overlap in space (graph vertices) in the absence of hard boundaries. This approach is therefore complementary to brain atlas that maps cortical areas into a set of discontinuous functional or structural regions (brain parcellations)[20–22]. A varied (and sometimes confusing) terminology has been used in relation to *connectivity gradients*, even though they are grounded on the same fundamental approach. If a comprehensive review of the growing gradients literature is out of the scope of this paper, Tables 1 and 2 aim at clarifying terminology and describing corresponding methodologies, while providing some key references.

Connectivity gradients estimated from functional[12], structural[23], or microstructural[24] brain data have been applied to study the hierarchical organization of brain structure and function. The pioneering work of Margulies and colleagues[12,25] introduced the concept of *gradients* to indicate eigenvectors obtained applying diffusion map embedding to resting-state fMRI connectivity matrix. Interestingly, the first non-trivial eigenvector of the spectral decomposition (corresponding to eigenvalue $\lambda_1$) revealed a macro-scale cortical organization spanning from unimodal sensory areas to transmodal association areas. Atasoy and colleagues[10] obtained *connectome harmonics* through Laplacian decomposition of DWI connectivity and showed that they can predict resting state functional networks activity. Results indicated that visual, sensory-motor and limbic networks more closely matched low-frequency harmonics while higher cognitive functions networks spanned over a broader range of brain modes. In[24], functional gradients were shown to only partially align with microstructural gradients obtained from histology and high-field MRI: if both axes of variance originates in primary sensory areas, the functional gradient identified in[12] arches towards default mode and fronto-parietal networks, while its microstructural counterpart extend to limbic cortices.

Earlier works have exploited Laplacian decomposition applied to the cortical surface mesh (instead of graphs of connectivity) to extract anatomically relevant features. In[26], Laplacian based spectral analysis was applied to the cortical mesh to investigate gyrification complexity. Using a segmentation of the cortex based on Laplacian spectral decomposition, the authors were able to identify developmentally relevant features such as primary, secondary and tertiary folds. Spatial eigenmodes of the curved cortex were also used in[27] to solve neural field corticothalamic equations and estimate so-called *activity eigenmodes* of the brain[28]. The authors also show that excitatory or inhibitory states can be reconstructed through a finite number of eigenfunctions and that these largely overlap with Laplacian eigenvectors of a connectivity matrix estimated from DWI imaging. Interestingly enough, and in line with our attempt at giving an unified view of the gradients and GSP methodologies, this approach linking neural field equations and Laplacian decomposition has been revisited in a GSP perspective in a recent publication[29] that introduce the Graph neural fields framework (see Table 1).

> **Box 1: Elements of Graph Theory and matrix representation**
>
> Graphs are tools that are ubiquitous in many fields of science, thanks to their genericity and expressivity. They allow to efficiently represent relations between items, called *vertices*. These relations are modeled using *edges*, which are most often *pairs* of vertices. For example, if $i$ and $j$ are vertices, $(i, j)$ is a potential edge in the graph. In many cases, relations are *weighted* with the convention that a weight 0 corresponds to the absence of an edge, and any nonzero value can be used otherwise. A concise way to represent a weighted graph consists in using its (weighted) adjacency matrix $W$, indexed by vertices. As such, $W_{ij}$ is 0 if and only if there is no edge between vertices $i$ and $j$, and $W_{ij}$ represents the weight of the edge $(i, j)$ otherwise.
>
> By summing a row of $W$, we obtain the *strength* (or weighted degree) of the corresponding vertex, which can be thought of as the importance it has, to be compared to the other vertices[30]. These strengths can be arranged in a diagonal matrix, called the strength matrix $D$, which allows to define the graph Laplacian: $\mathbf{L} = \mathbf{D} - \mathbf{W}$. For an undirected graphs with $N$ vertices, the Laplacian is a real symmetric matrix and thus has an orthogonal basis of eigenvectors $\mathbf{u}$ associated to eigenvalues $\lambda_l$ such as $\mathbf{Lu_l} = \lambda_l \mathbf{u_l}$, with $l = 0, 1, \cdots, N - 1$. The graph Laplacian is key to many fundamental properties about graphs, which can be found in the literature about spectral graph theory[31]. For example, providing $W$ contains only nonnegative values, the spectrum of $L$ is also nonnegative. It always contains the element 0. The magnitude of the second smallest eigenvalue of $\mathbf{L}$ $\lambda_1$ is an important indicator of the global *connectivity* of the graph: for example $\lambda_1 = 0$ if the graph is not connected (i.e. it is constituted of at least two separate subgraphs).
>
> In the field of Graph Signal Processing (GSP), authors are interested in manipulating graph matrices together with vectors indexed by the vertices of the graph, called *graph signals*. In the light of this formalism, it is possible to define ad-hoc graph Fourier modes, tied to the specific considered graph structure. These Fourier modes simply consist of the eigenvectors of $\mathbf{L}$. The corresponding eigenvalues $\lambda_l$ exhibit behaviors that can be interpreted in terms of spatial frequencies over the graph structure. In the case of simple ring graph structures, graph Fourier modes and classical discrete Fourier modes become identical. For more complex and arbitrary graphs, the abundant GSP literature explains how to design filters and other operators adapted to the underlying structure[6].
>
> In short, the graph Laplacian (and other versions of the Laplacian matrix, see Table 2) is ubiquitously used both in graph spectral analysis and GSP. In the first case, the properties of the Laplacian are typically exploited for dimensionality reduction of the graph (i.e. for the extraction of brain connectivity gradients from a brain network). In the case of GSP, Laplacian eigenvectors are used to decompose the signal in graph Fourier modes by defining a Graph Fourier Transform (see Table 2).
>
> It is worth clarifying that if these approaches are all grounded on a Laplacian-based spectral decomposition of a matrix, the adjacency matrix itself can be estimated with a variety of techniques (Figure 1, Panel a.). For instance, structural connectivity refers to anatomical connections between brain regions and is most commonly estimated using Diffusion Weighted Imaging (DWI). Functional connectivity is defined as the statistical dependence among measurements of neural activity[32] and it is usually inferred through correlations among neurophysiological time-series. Effective connectivity estimates the influence that one neuronal system exerts on an other and, because it refers to the notion of causality, is intrinsically directed. Functional and effective connectivity can be assessed by a range of neuroimaging techniques (Electroencephalography -EEG, functional Magnetic Resonance Imaging -fMRI, Magnetoencephalography -MEG). See also Tables 1 and 2 for more details on the terminology and methods of different approaches using Laplacian decomposition of brain signals.

Differently from the whole-brain approaches described hitherto, Haak and colleagues[33,34] proposed a framework to map regional connection topography (i.e. *connectopies*) using Laplacian eigenmaps[9]. For every voxel in the selected region, a connectivity fingerprint was obtained computing correlation with the fMRI time series from all the other gray matter voxels; non-linear manifold learning based on Laplacian decomposition of the connectivity matrix was then applied to extract corresponding eigenvectors named connectopies (i.e. connections topography). The authors underline the interest of using non-linear dimensionality reduction based on connectivity (instead of linear approaches such as PCA and ICA) to identify more biologically plausible maps of functional organization. Following this perspective, recent works have investigated the relation of connectivity gradients to cognitive processes, showing that these are altered depending on the ongoing cognitive experience or psychological state[13,35–38]. In[35], for example, macro-scale gradients of functional connectivity at rest are related to ongoing thoughts and different gradient profiles are associated to different cognitive task (i.e. problem-solving vs past-related thoughts).

Taken together, these studies underline the potential of data-driven dimensionality reduction based on brain networks to reveal principles of large-scale cortical organization and identify gradual changes in functional, white matter and cytoarchitectonic architecture in different conditions. They also suggest that connectivity gradients can yield meaningful functional relevance, and thus it might be particularly sensible to use them as a basis to analyze brain activity using GSP.

## 3 GSP for neuroimaging

The fundamental difference between GSP and graph theory is that while the latter provides tools to analyze and manipulate graphs, GSP focuses on analyzing signals *on* graphs. In other words, GSP leverages on concepts developed in spectral graph theory to translate Fourier analysis to signals on graphs. A general overview of the GSP framework illustrating how to perform operations on graphs (spectral analysis, convolution, filtering) is presented in[42] and[6] and a first review on GSP for neuroimaging was recently proposed[43]. Using a graph estimated from the white matter tracts of the brain, the authors detail how to apply graph Fourier transform to analyze brain activity, and exploit the corresponding graph frequency bands to interpret the data. Precisely, low frequencies correspond to brain activity that follows closely (i.e. is "aligned with") the underlying white matter connectivity, while high frequencies are characterized by brain activity patterns that can be seen as "liberal" with respect to the network structure.

This approach provides a simple and interpretable framework, that has successfully been applied to study the decoupling between brain structure and function in several recent works [17–19,44]. In[18] the application of GSP to model fMRI activity onto a DWI connectivity graph allowed a novel analysis of brain functional-structural coupling. The authors investigated to what extent fMRI time-series are constrained by the underlying structure. Results indicate that aligned signals concentrated within default mode and fronto-parietal systems, while subcortical system included both aligned and liberal modes. Interestingly, these findings are interpreted in terms of brain dynamics flexibility and linked to cognitive performance, showing that GSP can discriminate behaviorally relevant signals. Preti and colleagues[17] also decomposed resting-state fMRI time series using Laplacian eigenvectors of structural connectivity revealing a gradient of large-scale organisation of the cortex spanning from sensory-motor areas (with high functional-structural alignment) to higher cognitive functions areas (whose activity is more decoupled from underlying structure). In a seminal work by Glomb and colleagues[19], the GSP framework allowed for a sparser representation of source EEG data than the conventional individual regions analysis. Few structural connectivity harmonics were shown to capture EEG task dynamics and, more importantly, revealed significant patterns of activation involving the entire cortex, that were disregarded in the classical region-by-region analysis. Together with a high density EEG study in patients with disorder of consciousness[45], this is the only work applying GSP to the analysis of the EEG signals and it indicates that network harmonics also have a functional significance, as they can be considered as an orthogonal basis of large-scale EEG dynamics.

Going beyond metrics and inference-based approaches, other studies have combined GSP and machine learning to derive features from graph Fourier transform [46–49]. For instance in [47], a combination of GSP and machine learning was proposed for Autism Spectrum Disorder classification. More specifically, authors revealed that the analysis of fMRI data could be enriched by projecting resting state fMRI (rs-fMRI) time-series on a structural brain graph, as shown by substantial classification performance gains. In[49], the authors presented a feature extractor approach based on machine learning and spectral wavelets on brain graphs. In[50] a functional graph Laplacian embedding of deep neural networks (graph convolutional networks) is used to classify task fMRI time-series, in a joint GSP-deep learning framework. Finally, other approaches have taken advantage of graphs to denoise brain signals, such as in [51] where the authors simultaneously clean brain signals and learn the associated graph.

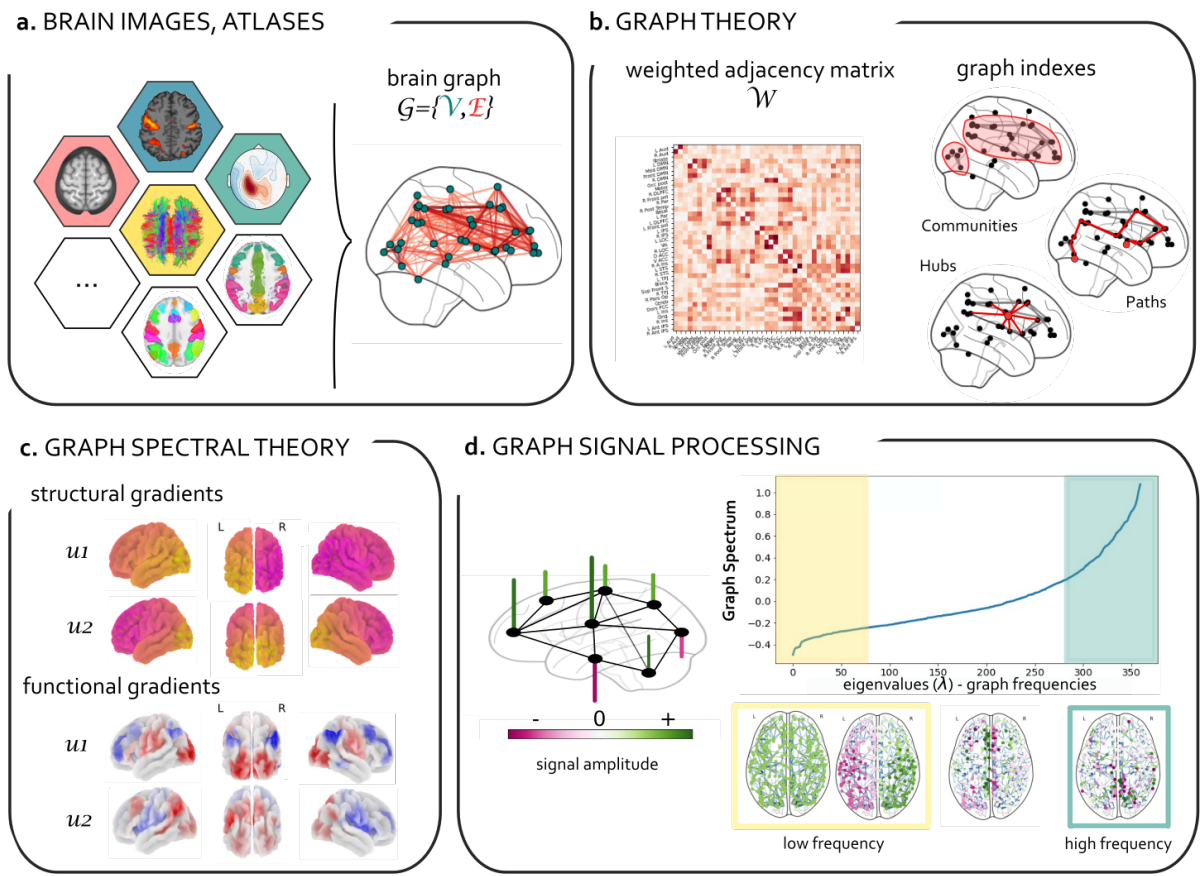

**Figure 1. From graph theory to graph signal processing in brain imaging. a.** Different areas of the brain can be represented as nodes and structural and functional relationship between them as edges of a complex large-scale network, also known as the *connectome*[39]. Various approaches exist to identify the nodes of the connectome (atlas and anatomic based, data-driven, etc.)[20, 21]. Similarly, edges of a brain network can be assessed with a range of neuroimaging techniques (DWI, EEG, fMRI, MEG, PET) and methods (structural, functional or effective connectivity[32]) **b.** Graph theory allows to describe salient properties of network topology with matrices (i.e. adjacency, Laplacian, Degree matrices, etc.) and graph indexes (i.e. efficiency, clustering, centrality)[1, 40, 41] **c.** Graph spectral analysis (e.g. Laplacian eigenvectors) is used to extract low dimensional representations of brain networks known as brain gradients[12] (see Tables 1 and 2). **d.** *Graph signal processing (GSP)*[6] takes a step forward as it associates a signal to an underlying graph. It extends classical analysis methods from regular domains (discrete time signals) to non-regular graphs. GSP allows to analyze brain activity taking into account the underlying topology of brain networks. GSP also allows for a spectral decomposition of brain activity based on the underlying graph Laplacian eigenvectors (Graph Fourier Transform, see Table 2). In the figure, a brain signal (whose amplitude is encoded in the height and color of the vertical bars) "lives" on an brain network (black) and can be decomposed in low (high) graph frequency harmonics corresponding to small (high) graph Laplacian eigenvalues.

> **Box 2: Toolboxes and Resources for Gradient analysis and GSP in Neuroscience**
>
> A variety of resources for graph analysis are available and extensively used in data science across multiple research domains. As graph structure and properties are encoded in matrices, in principle every toolbox manipulating arrays and matrices can be adapted to GSP and graph spectral theory; this includes for instance Python packages such as numpy, scikit, pytorch. Here we list toolboxes and Python modules that were specifically designed or that can be easily applied for GSP and graph spectral analysis of brain networks and their visualization.
>
> - **Nilearn**: a widely used Python toolbox for machine learning in neuroimaging. It also includes useful functions for brain connectivity computation and visualization that can easily adapted to plot gradients and signals on brain networks. https://nilearn.github.io/
> - **BrainSpace**: a Matlab/Python software package that allows gradients computation and analysis specifically adapted to neuroimaging and connectome datasets[52]. https://github.com/MICA-MNI/BrainSpace
> - **PyGSP**: a Python package specifically designed for graph signal processing that implements a variety of operations on graphs (computing Graph Fourier transform, filtering or interpolating signals on graphs, plotting) that also scale to very large graphs.
>   https://pygsp.readthedocs.io/en/stable/index.html
> - **NetworkX**: a Python package to analyze network structure, build network models, network algorithms and visualize networks. https://networkx.github.io/
> - **Congrads**: a Python package compute and map connectopies (connectivity topography) of a pre-defined region-of-interest. https://github.com/koenhaak/congrads

## 4 Discussion and perspectives

In this paper we provided an overview of works that applied Laplacian spectral decomposition and GSP to analyze brain signals. We attempted to elucidate terminology and related approaches used in the "gradients" and GSP communities, systematically describing recent promising results. In this section we discuss the potential of an integrated gradients/GSP framework to reveal a spectral basis of brain activity grounded on brain connectivity topology. We will also bring up open questions and methodological challenges of this novel approach.

### 4.1 Connectivity Gradients as a Fourier basis of brain activity

Recent works applying Laplacian-based spectral decomposition of brain networks have revealed functionally, developmentally and anatomically relevant hierarchies of organization of the brain that have also been related to cognitive performance. Connectivity gradients obtained by such graph spectral decomposition not only have been shown to represent relevant axis of brain organization, but within the GSP framework can be seen as a Fourier basis to decompose brain activity. In this sense they provide a new spatial-frequency language to characterize patterns of neural activity and a novel perspective of probing brain dynamics.

A few studies have investigated the relationship between the principal[36–38] or second[35] gradient of rs-fMRI connectivity during various tasks, suggesting that the cortical organization encoded in connectivity gradients support specific cognitive or semantic functions. The GSP framework allows to take the analysis further by using the whole spectrum of connectivity gradients as a Fourier basis to decompose brain activity. For instance, the large-scale gradients identified in[12] could be exploited to disentangle brain activity measured during a complex cognitive task in unimodal (e.g. related to sensory processing) and transmodal (e.g. related to ongoing thoughts) patterns. This analysis would extend the work of[38] relating local and distributed processes to the organization of the cortex in unimodal and transmodal areas. Moreover, by decomposing brain signals as a function of eigenvectors (or gradients) of the underlying connectivity, the GSP framework allows exploiting the information encoded in higher order connectivity gradients whose functional relevance has been scarcely explored in literature. For instance in[17] the spectrum of structural connectivity eigenvectors is split in low and high frequency components to define a binary decoupling index (low frequencies correspond to signals coupled to structural connectivity, while high frequency components are considered as decoupled). The potential of the GSP framework could be further exploited by considering the whole set of connectivity gradients, instead of a partition in low vs high frequencies, in a similar way as classical Fourier analysis is used to decompose signals in the time domain. Continuing in the analogy with the classical frequency domain, fundamental operations such as filtering and denoising can be generalized to brain signals on graph by taking in account the full graph spectrum. For instance, artifactual components of brain activity (i.e. balistocardiogram artifact for EEG-fMRI simultaneous acquisition[53,54]) could be reduced by filtering out the graph frequency component or band that best represents the artifact (i.e. that maximally correlates with electrocardiogram signal).

Using a continuous set of dimensions (graph Fourier modes) for the analysis of brain dynamics is an approach that complements (rather than exclude) the more classical hard boundaries parcellation. Some processes may be best characterized in terms of non-overlapping fixed regions, others in terms of delocalized, overlapping eigenmodes. In this sense, this graph modal approach may be more appropriate than modular analysis in describing complex cognitive states depending on multiple overlapping phenomena. Neural patterns of ongoing activity could be seen as location in a multidimensional state-space constructed out of large-scale gradients[12]. The "biological" validity of this approach can be also found in the intrinsic organization of brain tissues. Brain structure (and function) are organized in overlapping hierarchical components[55]. It is well known, for instance, that the visual and auditory cortices are organized in topographic maps that reflects how sensory information is processed (i.e. retinotopic or tonotopic mapping[56]). As a result, in the same cortical area multiple and heterogeneous modes coexist. Moreover, while the cytoarchitecture of a region can be considered as uniform, the same region can be heterogeneous in terms of function, gene profile or axonal projections (i.e. connectivity topography[33]). This intrinsic complexity may be better represented by a continuum of functions instead of a mosaic of brain areas[57].

The GSP approach also allows to efficiently integrate different neuroimaging techniques (EEG, fMRI, DWI, MEG) thus exploiting complementary measurements of brain properties. Only few works have analyzed fMRI and EEG signals at rest using graph Fourier modes of the underlying structural graph (i.e. estimated form DWI). In the future, this framework could be applied to jointly decompose electrophysiological and functional time-series using underlying structural topology, thus integrating different temporal and spatial scales in a multimodal analysis, with potential to shed new light on the complex interplay between function and structure of the brain. In addition, GSP could be extended to the analysis of brain signals during different tasks and (pathological) conditions, holding a promise for developing more sensitive markers of disease.

**4.2 Methodological challenges and Future Directions**

As described in Table 2, even if grounded on the same fundamental approach, various algorithms have been applied for graph spectral analysis. These methodologies have different properties and it is not clear which one should be used and for which specific applications. For instance, different Laplacian matrices can be used for spectral decomposition. The normalized Laplacian has the useful property that its spectrum is limited to the $[0, 2]$ interval however its first eigenvector (associated with $\lambda_0 = 0$) is not constant as for the combinatorial Laplacian: this is less intuitive for an interpretation of eigenvalues in terms of spatial frequency[6]. Similarly, influential studies used Diffusion Maps embedding[12,35,58] with specific parameters choice (diffusion time and anisotropic diffusion parameter $\alpha$) because of a series of advantageous properties for brain connectivity analysis. Future works should explore how different choices could affect results and which metrics are more adapted to the analysis of specific graph structures or brain signals.

Another point to clarify is the impact of the choice of the parcellation on the connectivity gradients topography. If in the pioneering work of[12] high resolution rs-fMRI connectivity was computed between fMRI voxels, in other works[17,35,59] different atlas were used to identify brain ROIs and then compute functional connectivity between them. It would be interesting to assess how the use of different parcellations affect the computation of connectivity gradients and if it introduces some bias in the characterization of connectivity topography, that are supposed to describe a continuous patterns of organization.

One important limitation of the works we have reviewed is the lack of accounting for temporal dependencies in the graph model. Indeed, in bulk of the literature "constant" spatial dependencies between brain regions are considered (i.e. structural graphs built using white matter tracts, or functional graphs estimated using statistics of brain activity). Several theoretical progresses have been made to address dynamic (i.e. time-varying graphs), and their application to neuroimaging data could prompt the understanding of cognitive processes dynamics. Some promising frameworks to model time-varying aspects in graphs include graph slepians[60], sparseness of temporal variation[61] or lapped Fourier transform[62]. Another avenue to model spatio-temporal dynamics could be to use deep learning models adapted to sequence modeling, combined with graph convolutional networks[63].

Finally, GSP is an active research field and there are a few recent theoretical proposals in GSP that have not yet been applied to neuroimaging data, but could potentially bring interesting breakthroughs. By considering generalized signal processing operations on graphs such as graph filters[64], graph wavelets[65,66], multiscale graphs[67], graph slepians[43], graph sampling[68,69] or locating and decomposing signals on graphs[70] could enable richer interpretations, and potentially a unified perspective on graph signals.

In conclusion, this work reviewed recent studies applying gradients and GSP for the analysis of brain signals, clarified terminology and methods and related these two approaches grounded on the eigenvectors decomposition of connectivity matrix. We point out that, given the increasing relevance connectivity gradients are taking in the understanding of brain macro-scale organization, the application of GSP to neuroimaging is an exciting avenue towards a deeper understanding of brain organization. We also identify methodological challenges and suggest that future works should address multimodal and time-varying modelisation and further explore the use of different metrics.

| | Approach | Toolbox | Key references |
|---|---|---|---|
| CONNECTOME HARMONICS, HARMONIC BRAIN MODES | Laplacian eigenvectors $u_i$ of large-scale brain networks estimated from DWI and anatomical MRI. They are interpreted as spectral components of spatio-temporal neural activity and compared to resting state networks and oscillatory patterns (Neural Field model). | | 10, 13, 71 |
| BRAIN GRADIENTS | Eigenvectors $u_i$ obtained applying Diffusion Map embedding on large-scale functional[12], microstructural[58] or spontaneous oscillations[72] networks estimated respectively from resting-state fMRI data, myelin-sensitive MRI data or MEG signals. They reveal macro-scale axes of cortical organization with functional and neurodevelopment relevance. | BrainSpace | 12, 24, 35, 73, 74 |
| CONNECTOPIES | Laplacian eigenvectors $u_i$ of the graph obtained computing the correlation between voxels within a selected ROI and the rest of gray matter voxels. The approach reveals fine-grained topographic organization of a brain region's connectivity (i.e. primary motor or visual cortex). | | 33, 34 |
| BRAIN ACTIVITY EIGENMODES | Excitatory or inhibitory neural activity expanded in terms of spatial eigenmodes of the cortex mesh obtained solving corticothalamic Neural Field Theory equations. These brain eigenmodes show high similarity with spherical harmonics (cortical folding=0) and DWI connectivity eigenvectors (graph Laplacian). | Congrads | 27 |
| FOURIER, HARMONIC MODES | Graph Fourier modes obtained applying graph Fourier transform to a signal (i.e. fMRI[11,75] or EEG[19]) on a graph (i.e. structural connectivity graph estimated from DWI). This analysis reveals low (high) frequency modes that are *aligned* (*liberal*) with respect to the underlying graph structure. | PyGSP | 16–19 |
| GRAPH NEURAL FIELDS | Excitatory or inhibitory neural activity expressed as stochastic neural field equations on the human connectome graph. This approach combine Wilson-Cowan neural field equations and graph signal processing to to model and analyze whole-brain activity. | Code available [†] | 29 |
| Notes | [†] https://github.com/marcoaqil/Graph-Stochastic-Wilson-Cowan-Model | | |

**Table 1.** Graph spectral analysis and GSP applied to Neuroscience: **terminology**. First 3 rows: based on spectral graph theory (spectral decomposition of brain networks). Last two rows (Light blue): using GSP (Graph Fourier Transform: spectral decomposition of a brain signal based on the underlying brain graph topology).

| | Notation | Method | Key references |
|---|---|---|---|
| LAPLACIAN EMBEDDING or EIGENMAPS | *Operator* <br> $L = D - W$ Laplacian <br> $L_n = D^{-1/2}LD^{-1/2}$ Normalized Laplacian <br> $L_{rw} = D^{-1}L$ Random Walk Laplacian <br><br> *Low-dimensional Representation* <br> $G = [u_1, u_2, u_3, \ldots]$ | Spectral decomposition of a graph $G$ in eigenvectors of the graph Laplacian $L$. It is the discrete counterpart (on graph) of the Laplacian-Beltrami operator on continuous manifolds. The Laplacian eigenvectors associated with the lowest eigenvalues provide a dimensionality reduction mapping that preserves locality. | 33,34 **9**, |
| DIFFUSION MAPS | *Operator* <br> $W(\alpha) = D^{-1/\alpha}WD^{-1/\alpha}$ <br> $P(\alpha) = D(\alpha)^{-1}W(\alpha)$ Transition Probability <br><br> *Low-dimensional Representation* <br> $G = [\lambda_1^T u_1, \lambda_2^T u_2, \ldots]$ | Diffusion map embedding treats the graph $G$ as the basis of a diffusion process. Diffusion maps are a family of graph Laplacians that depends on a diffusion parameter $\alpha$. They can be employed to embed the data into a Euclidean space where the probability of transition between nodes defines the Euclidean distance between the corresponding points in the embedding space. | 12,25,35,73,76 **77,78**, |
| GRAPH SIGNAL PROCESSING | *Laplacian Eigendecomposition* <br> $L = U\Lambda U^T$ <br><br> *Graph Fourier Transform of X* <br> $\tilde{X} = U^T X$ | Expansion of a signal (or a stochastic function) $X$ in terms of the eigenvectors of the underlying graph Laplacian $L$. Laplacian eigenvectors carry a notion of spatial frequency: i.e. eigenvectors corresponding to low eigenvalues vary smoothly across the graph; those corresponding to large eigenvalues have higher spatial frequencies (i.e. are more likely to have different values across adjacent vertices). | 16–19,29 **6,42**, |

**Table 2.** Graph spectral analysis and GSP applied to Neuroscience: **methods**. For each methodology, a brief description of the approach and main references are provided (in bold the key reference describing in detail the algorithm). First two rows: based on spectral graph theory (spectral decomposition of brain connectivity graphs). Last row (Light blue): using GSP (Graph Fourier Transform: spectral decomposition of a brain signal based on the underlying brain graph topology). *Notations*: *W*, graph adjacency matrix; *D*, degree matrix; $\lambda$, eigenvalues and **u**, eigenvectors of the embedding operator. $\alpha$ diffusion operator; $\Lambda$, diagonal matrix of eigenvalues; U, matrix whose columns are the eigenvectors $u_i$.